\pdfoutput=1
\documentclass[twocolumn,amsmath,nobibnotes,prb,aps]{revtex4-1}
\usepackage{bm}
\usepackage{graphicx}

\newcommand{\TCT}{{T_\mathrm{C \leftrightarrow T}}}
\newcommand{\TTO}{{T_\mathrm{T \leftrightarrow O}}}
\newcommand{\TOR}{{T_\mathrm{O \leftrightarrow R}}}

\begin{document}
\title{Molecular Dynamics Simulations of Chemically Disordered Ferroelectric
(Ba,Sr)TiO$_3$ with a Semi-Empirical Effective Hamiltonian}
\author{Takeshi Nishimatsu$^1$}
\author{Anna Gr\"unebohm$^2$}
\author{Umesh V. Waghmare$^3$}
\author{Momoji Kubo$^1$}
\affiliation{
$^1$Institute for Materials Research (IMR), Tohoku University, Sendai 980-8577, Japan\\
$^2$Faculty of Physics and Center for Nanointegration, CENIDE,
University of Duisburg-Essen, 47048 Duisburg, Germany\\
$^3$Theoretical Sciences Unit, Jawaharlal Nehru Centre for Advanced Scientific Research (JNCASR),
Bangalore-560064, India}

\begin{abstract}
We present a semi-empirical effective Hamiltonian to capture effects of disorder associated
with Ba and Sr cations occupying $A$ sites in (Ba$_{x}$Sr$_{1-x}$)TiO$_3$ on its ferroelectric phase transition.
Averaging between the parameters of first-principles effective Hamiltonians of end members
BaTiO$_3$ and SrTiO$_3$, we include a term with an empirical parameter to capture the local polarization and
strains arising from the difference between ionic radii of Ba and Sr.
Using mixed-space molecular dynamics of the effective Hamiltonian, we determine $T$-dependent ferroelectric
phase transitions in (Ba$_{x}$Sr$_{1-x}$)TiO$_3$ which are in good agreement with experiment.
Our scheme of determination of semi-empirical parameters
in effective Hamiltonian should be applicable to other perovskite-type ferroelectric solid solutions.
%Parameters of an effective Hamiltonian for molecular-dynamics (MD) simulation of
%barium-strontium titanate solid solution, (Ba$_{x}$Sr$_{1-x}$)TiO$_3$, are determined semi-empirically.
%The parameters are given as averages of those for BaTiO$_3$ and SrTiO$_3$.
%Change in lattice constants are effectively given through $x$-dependent external hydrostatic pressure.
%Effect of \textit{modulation} on local inhomogeneous strains caused by
%the difference in ionic radii of Ba and Sr is clarified.
%Procedure of first-principles calculations, determination of parameters, and MD calculations
%described here may be applicable to other perovskite-type ferroelectric solid solutions.
\end{abstract}

%\pacs{77.80.Dj, 77.80.Fm, 64.70.Nd}
% 77.80.Dj  Domain structure; hysteresis
% 77.80.Fm  Switching phenomena
% 64.70.Nd  Structural transitions in nanoscale materials

\maketitle

\section{Introduction}
Barium (Ba) and strontium (Sr) belong to the same column of the periodic table
and are chemically very similar.
However,
ternary perovskites $AB$O$_3$ based on Ba and Sr at their $A$ site can be quite different
in their properties:
barium    titanate (BaTiO$_3$) is ferroelectric at room temperature, while
strontium titanate (SrTiO$_3$) is  paraelectric.
It is thought that this is largely because of the difference in
their ionic radii\cite{ShannonEffectiveIonicRadii},
$r_\mathrm{Ba}=1.61$~\AA\ %
and $r_\mathrm{Sr}=1.44$~\AA.
It can be more understandable through
the tolerance factor~\cite{Goldschmidt:ToleranceFactor} defined as
\begin{equation}
  \label{eq:ToleranceFactor}
  t=\frac{r_{A}+r_\mathrm{O}}{\sqrt{2}(r_{B}+r_\mathrm{O})}\ ,
\end{equation}
where, $r_A$, $r_B$, and $r_\mathrm{O}$ are
ionic radii of cation $A=\mathrm{Ba}^{2+}$ or Sr$^{2+}$,
cation $B=\mathrm{Ti}^{4+}$, and O$^{2-}$, respectively.
$t=1.062$ for BaTiO$_3$ means that
the $B$-site Ti ion is too small for its site,
the ion can shift off-centeringly, leading to the occurrence
of displacive-type ferroelectricity in the crystal\cite{Fu:Itoh:Silver.Perovskite.Oxides:2011}.
$t$ is almost unity ($t=1.002$) for SrTiO$_3$, indicating that there is no room for ions to move,
ideal cubic perovskite structure become stable at room temperature,
and indeed SrTiO$_3$ does not show ferroelectricity
down to the absolute 0~K.
% Ba Ti+4 O  1.610, 0.605, 1.400,   t=1.062
% Sr Ti+4 O  1.440, 0.605, 1.400,   t=1.002
Experimentally,
at low temperature ($T<106$~K), SrTiO$_3$ exhibits
very small rotational shift of oxygen octahedra ($\approx 1.6^\circ$)
and results in antiferrodistortive \textit{I4/mcm} structure\cite{doi:10.1143/JPSJ.46.581}.
At very low temperatures, intrinsic quantum paraelectricity\cite{MULLER:B:PRB:19:p3593-3602:1979}
is also found in SrTiO$_3$.

It has been found experimentally\cite{PhysRevB.54.3151.BST.Lemanov,PhysRevB.65.224104} that
the three transition temperatures of BaTiO$_3$,
cubic        $\leftrightarrow$ tetragonal $\TCT$,
tetragonal   $\leftrightarrow$ orthorhombic $\TTO$, and
orthorhombic $\leftrightarrow$ rhombohedral $\TOR$
decrease almost linearly, when Ba composition $x$ of
(Ba$_{x}$Sr$_{1-x}$)TiO$_3$ is reduced from 1.
Around pure SrTiO$_3$ ($x < 0.094$), it is known that the solid solution becomes almost cubic,
or more precisely,
antiferrodistortive \textit{I4/mcm} structure with very small atomic displacements.

%\textbf{NEWLY ADDED (1):}
Perovskite-type ferroelectric solid solutions such as (Ba$_{x}$Sr$_{1-x}$)TiO$_3$
are of great interest in the field of dielectrics, since many commercial
high-dielectric-constant material structures consist of such solid solutions\cite{McQuarrie:JACE:v38:p444:y1955}
and the composition parameter ($x$ here) is adjusted to get desired properties.
Therefore, offering a recipe of computational simulations of such solid solutions is important.

In 2006,
Walizer \textit{et al.} presented\cite{Bellaiche.PhysRevB.73.144105}
Monte Carlo simulations with an effective Hamiltonian
determined from first-principles calculations of (Ba$_{1/2}$Sr$_{1/2}$)TiO$_3$
within a virtual crystal approximation (VCA) and
local inhomogeneous strains determined from
fully disordered ionic configurations of Ba and Sr of (Ba$_{x}$Sr$_{1-x}$)TiO$_3$.
They successfully reproduced
the temperature--composition ($T$--$x$) phase diagram,
though with a large underestimation of polarization.
That underestimation basically came from
a local-density approximation (LDA).
Moreover,
because local inhomogeneous strains around each site
were fixed in their analysis and
were not allowed to fluctuate thermally,
temperature dependence of
the effect from ionic configuration
was not so clear.

Here, we newly determine a set of parameters for an effective Hamiltonian
for (Ba$_{x}$Sr$_{1-x}$)TiO$_3$ from more accurate first-principles calculations,
and perform molecular-dynamics (MD) simulations.
In our MD simulations,
local inhomogeneous strains around each site are not fixed but can fluctuate thermally.
We report not only a temperature--composition ($T$--$x$) phase diagram,
but also the dependence of polarization and lattice constants on composition.

In Sec.~\ref{sec:Formalism},
we briefly describe the first-principles methods we employ
and the formalism and conditions of our MD simulations.
In Sec.~\ref{sec:results}, we present results of our MD simulations,
and finally summarize our work in Sec.~\ref{sec:summary}.

\section{Methods of Calculation and Formalism}
\label{sec:Formalism}
\subsection{First-Principles Methods}
Our first-principles calculations are based on the density functional theory (DFT) as implemented in
ABINIT code\cite{Gonze:ABINIT.ComputMaterSci:2002,ABINIT_CPC_2009,ABINIT_CPC_2016}.
Bloch wave functions of electrons are expanded in the
plane wave basis truncated with a cut-off energy of 60~Hartree,
and are sampled on an
$8\!\times\! 8\!\times\! 8$ grid of $k$-points
in the first Brillouin zone.
We do not use LDA but use ``Wu and Cohen''\cite{Wu:C:PRB:73:p235116:2006} GGA functional,
along with Rappe's optimized pseudopotentials\cite{RAPPE:R:K:J:PRB:41:p1227-1230:1990}
generated with Opium code\cite{opium}.
A valley-line tracing method\cite{Hashimoto:N:M:K:S:I:JJAP:43:p6785-6792:2004}
is used to determine total energy surface of BaTiO$_3$ and SrTiO$_3$.
We basically use results of first-principles calculations of BaTiO$_3$ and SrTiO$_3$
in Ref.~\onlinecite{Nishimatsu.PhysRevB.82.134106}.

\subsection{Effective Hamiltonian}
\label{sub:Effective:Hamiltonian}
We use an effective Hamiltonian,
obtained with input from first-principles calculations,
for MD simulations.
It is essentially the same as that in
Refs.~\onlinecite{Nishimatsu.PhysRevB.82.134106}~and~\onlinecite{Nishimatsu:feram:PRB2008},
\begin{multline}
  \label{eq:Effective:Hamiltonian}
  H^\mathrm{eff}%(\{\bm{u}\},\{\bm{w}\}, \eta_1,\cdots\!,\eta_6)
  = \frac{M^*_\mathrm{dipole}}{2} \sum_{\bm{R},\alpha}\dot{u}_\alpha^2(\bm{R})
  + \frac{M^*_\mathrm{acoustic}}{2}\sum_{\bm{R},\alpha}\dot{w}_\alpha^2(\bm{R})\\
  + V^\mathrm{self}(\{\bm{u}\})+V^\mathrm{dpl}(\{\bm{u}\})+V^\mathrm{short}(\{\bm{u}\})\\
  + V^\mathrm{elas,\,homo}(\eta_1,\cdots\!,\eta_6)+V^\mathrm{elas,\,inho}(\{\bm{w}\})\\
  + V^\mathrm{coup,\,homo}(\{\bm{u}\}, \eta_1,\cdots\!,\eta_6)+V^\mathrm{coup,\,inho}(\{\bm{u}\}, \{\bm{w}\})~,
\end{multline}
where the phase space of atomic motion is reduced to a subspace spanned by
local      soft     mode    vectors $\bm{u}(\bm{R})$ and
local acoustic displacement vectors $\bm{w}(\bm{R})$
of each unit cell at $\bm{R}$ in a simulation supercell.
$\eta_1,\dots,\eta_6$ are the six components of homogeneous strain in Voigt notation ($\eta_1=e_{xx}$, $\eta_4=e_{yz}$).
$\frac{M^*_\mathrm{dipole}}{2} \sum_{\bm{R},\alpha}\dot{u}_\alpha^2(\bm{R})$ and
$\frac{M^*_\mathrm{acoustic}}{2}\sum_{\bm{R},\alpha}\dot{w}_\alpha^2(\bm{R})$ are the
kinetic energies of local soft modes and
local acoustic displacements along with their effective masses of $M^*_\mathrm{dipole}$ and $M^*_\mathrm{acoustic}$,
$V^\mathrm{self}(\{\bm{u}\})$ is the local-mode self-energy,
$V^\mathrm{dpl}(\{\bm{u}\})$ is the long-range dipole-dipole interaction,
$V^\mathrm{short}(\{\bm{u}\})$ is the short-range harmonic interaction between local soft modes,
$V^\mathrm{elas,\,homo}(\eta_1,\dots,\eta_6)$ is the elastic energy from homogeneous strains,
$V^\mathrm{elas,\,inho}(\{\bm{w}\})$ is the elastic energy from inhomogeneous strains,
$V^\mathrm{coup,\,homo}(\{\bm{u}\}, \eta_1,\dots,\eta_6)$ is the coupling between the local soft modes and the homogeneous strain, and
$V^\mathrm{coup,\,inho}(\{\bm{u}\}, \{\bm{w}\})$ is the coupling between the soft modes and
the inhomogeneous strains.
Detailed explanation of symbols in the effective Hamiltonian
can be found in
Refs.~\onlinecite{Nishimatsu:feram:PRB2008},
\onlinecite{King-Smith:V:1994}, and
\onlinecite{Zhong:V:R:PRB:v52:p6301:1995}.
To decrease the computational time,
forces exerted on $\{\bm{u}\}$ are calculated
in reciprocal space
using fast-Fourier transform (FFT) methods\cite{Waghmare:C:B:2003,Nishimatsu:feram:PRB2008,ACR:Waghmare:2014}.

% Hydrostatic pressure $p$ is applied to the system trough
% the homogeneous elastic energy  $V^{\rm elas,\,homo}(\eta_1,\cdots\!,\eta_6)$ as
% \begin{eqnarray}
%   \label{eq:V:elas:homo}
%   \nonumber
%   V^{\rm elas,\,homo}(\eta_1,\cdots\!,\eta_6)
%   & = & \frac{N}{2}B_{11}(\eta_1^2+\eta_2^2+\eta_3^2)\\
%   \nonumber
%   & + & N          B_{12}(\eta_2\eta_3+\eta_3\eta_1+\eta_1\eta_2)\\
%   & + & \frac{N}{2}B_{44}(\eta_4^2+\eta_5^2+\eta_6^2)~,
% \end{eqnarray}

\subsection{Effects of $A$-site Ordering with Ba or Sr Ions}
\label{sub:modulation}
\begin{figure}
  \centering
  \includegraphics[width=70mm]{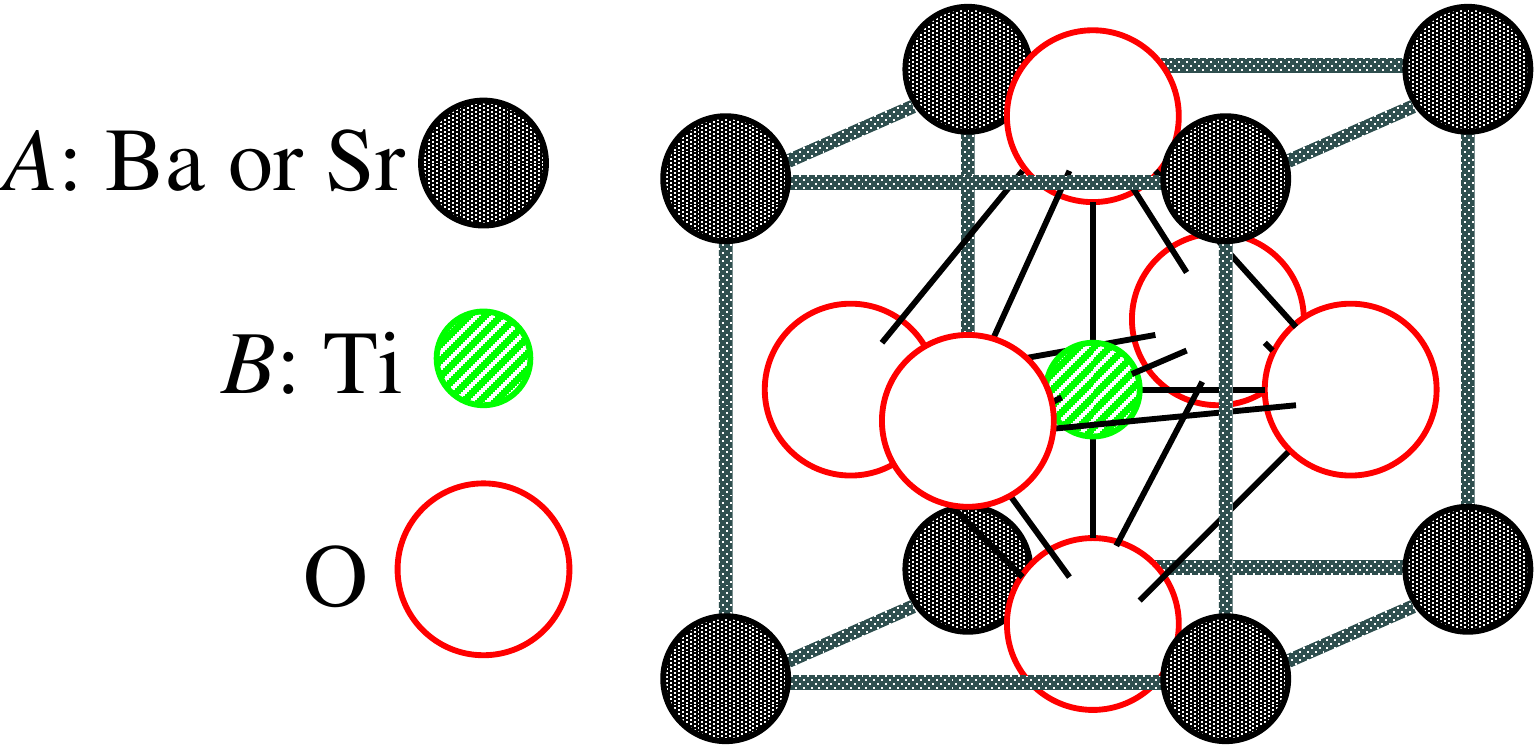}
  \caption{(Color online) Schematic illustration of
    perovskite-type crystal structure of (Ba,Sr)TiO$_3$.
    $s(\bm{R})=-8, -6, -4, -2, 0, +2, +4, +6, +8$ of Eq.~(\ref{eq:count})
    is the number of Ba ($+1$) or Sr ($-1$) ions at eight $A$-sites
    surrounding a given $B$-site at $\bm{R}$.}
  \label{fig:ABO3}
\end{figure}
To include the effects of alloying Ba and Sr with different ionic radii,
according to Ref.~\onlinecite{Bellaiche.PhysRevB.73.144105},
we count the number of
Ba or Sr ions at the 8 $A$-sites surrounding a given $B$-site at $\bm{R}$,
\begin{equation}
  \label{eq:count}
  s(\bm{R}) = \sum_{i=1}^8 \sigma_i
\end{equation}
as shown in Fig.~\ref{fig:ABO3},
where $\sigma_i = +1$ or $-1$ corresponds to the presence of a Ba or Sr ion, respectively.
Consequently, value of $s(\bm{R})$ ranges $-8, -6, -4, -2, 0, +2, +4, +6, +8$.
In contrast to Ref.~\onlinecite{Bellaiche.PhysRevB.73.144105},
we introduce a term for \textit{modulation} in local inhomogeneous strains by adding
\begin{multline}
  \label{eq:modulation}
    V^\mathrm{modulation,inho}(\{\bm{w}\},\{s\})\\
    =       c     \sum_{\bm{R}} \sum_{\alpha=1,2,3} s(\bm{R}) \, \eta_\alpha(\bm{R})\\
    = \frac{c}{N} \sum_{\bm{k}} \sum_{\alpha=x,y,z} \widetilde{w}_\alpha^\ast(\bm{k}) \, k_\alpha \, \widetilde{s}(\bm{k})
\end{multline}
to $H^\mathrm{eff}$ of Eq.~(\ref{eq:Effective:Hamiltonian}).
Here,
$c$ is strength of the \textit{modulation},
$N$ is the number of unit cells in the supercell,
$\bm{k}$ is wave vector,
$\widetilde{w}_\alpha^\ast(\bm{k})$ is complex conjugate of Fourier transform of $\bm{w}(\bm{R})$,
$\widetilde{s}(\bm{k})$ is Fourier transform of $s(\bm{R})$.
To simplify the computation,
$V^\mathrm{modulation,inho}(\{\bm{w}\},\{s\})$ is calculated in reciprocal space.

Effective hydrostatic pressure of
\begin{equation}
  \label{eq:effective:p}
  p = b(\frac{1}{2}-x)
\end{equation}
is applied to capture the
homogeneous strain that alters the lattice constants as a function of $x$,
because the \textit{modulation} of Eq.~(\ref{eq:modulation}) does not
include alternation of the homogeneous strain.
Here, $b$ is a constant.
Temperature-dependent negative effective pressure $p=-0.005T$~GPa for BaTiO$_3$,
which was applied in Ref.~\onlinecite{Nishimatsu.PhysRevB.82.134106}
to mimic thermal expansion,
is not applied in present work.

%\textbf{NEWLY ADDED (2):}
In the present MD simulations, only the parameters $V^\mathrm{modulation,inho}(\{\bm{w}\},\{s\})$ and
$p$ are $x$-dependent,
and other parameters in the effective Hamiltonian of Eq.~(\ref{eq:Effective:Hamiltonian})
are kept constant.
Such simplification can be successfully employed because
Ba and Sr are chemically very similar and different only in their ionic radii.
Determination and exact values of the parameters $c$ and $b$ will be discussed in Sec.~\ref{subsec:alloying}.

\subsection{Molecular-Dynamics (MD) Simulations}
\label{subsec:MD:conditions}
MD simulations of (Ba$_{x}$Sr$_{1-x}$)TiO$_3$ ($x=0.0$, $0.1$, $\dots$, $0.9$, $1.0$)
with the effective Hamiltonian
% of Eq.~(\ref{eq:Effective:Hamiltonian})
are performed with our original MD code {\tt feram}\cite{feram}.
Input files for present simulations are in its free software package of
{\tt feram-0.26.01/src/34example-BST/}, and
details of the code can be found in Ref.~\onlinecite{Nishimatsu:feram:PRB2008}.
Temperature is kept constant in each temperature step
of simulation within a canonical ensemble
using the velocity-scaling thermostat for both $\{\bm{u}\}$ and $\{\bm{w}\}$ with the
time step of $\Delta t=2$~fs.
We use a supercell with size of $N = L_x\times L_y\times L_z = 32 \times 32 \times 32$ unit cells
and temperature steps of $\pm 1$~K/step in heating-up and cooling-down simulations.
In every temperature step,
we thermalize the system for 20,000 time steps,
after which we use 20,000 time steps to average the properties.
The initial configurations of $\{\bm{u}\}$ are generated randomly:
$\langle u_\alpha \rangle = 0.11$\AA\ ($\alpha=x,y,z$) for heating-up simulations,
$\langle u_\alpha \rangle = 0.00$\AA\ for cooling-down simulations, and
variance of $\langle u_\alpha^2 \rangle - \langle u_\alpha \rangle^2=$~(0.02~\AA)$^2$ for the both.
In the initial configurations, $\{\bm{w}\}$ are set to zero.
We have checked that the results of these simulations
do not depend on initial configurations.
A set of $s(\bm{R})$ for each $x$ of
(Ba$_{x}$Sr$_{1-x}$)TiO$_3$
is generated from random configurations of
$xN$ Ba and $(x-1)N$ Sr ions.
%It is in this aspect that our simulations go beyond the virutal crystal approximation.

\section{Results and Discussion}
\label{sec:results}

\subsection{Results of First-principles Calculation and
  Determination of Parameters of $H^\mathrm{eff}$ of (Ba$_{1/2}$Sr$_{1/2}$)TiO$_3$}
\label{subsec:determination}
\begin{figure}
  \centering
  \includegraphics[width=55mm]{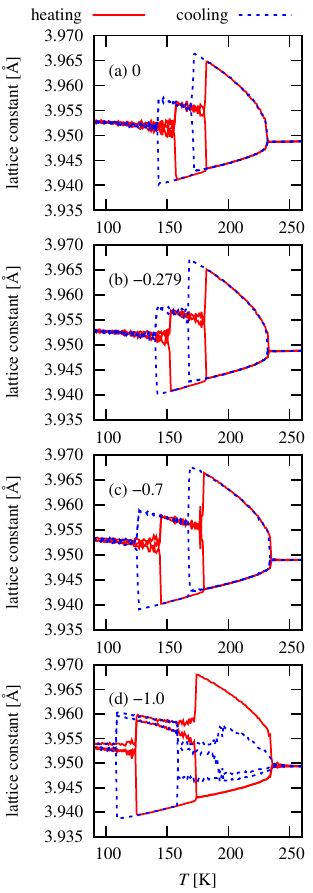}
  \caption{(Color online) Simulated temperature-dependence of lattice constants of
    (Ba$_{1/2}$Sr$_{1/2}$)TiO$_3$ for four different strength of the \textit{modulation},
    $c = 0, -0.279, -0.7, -1.0$, (a)--(d), respectively. The larger $|c|$, the lower $\TOR$.
    %The other two transition temperatures $\TCT$ and $\TTO$ are relatively unchanged.
  }
  \label{fig:half}
  % app2:/home/t-nissie/feram/feram-q.00/xBST/xBST-08-c-dep/c-dep.gp
\end{figure}

Using the systematic procedure described in
Ref.~\onlinecite{Nishimatsu.PhysRevB.82.134106},
we perform first-principles calculations to determine
a set of parameters of $H^\mathrm{eff}$ for SrTiO$_3$ (See Table~\ref{tab:parameters}).
We averaged the parameters of $H^\mathrm{eff}$ of BaTiO$_3$ in Ref.~\onlinecite{Nishimatsu.PhysRevB.82.134106} and
those of $H^\mathrm{eff}$ of SrTiO$_3$ (See Table~\ref{tab:parameters}).
It is found that this set of parameters indeed reproduces the three transition temperatures of
(Ba$_{1/2}$Sr$_{1/2}$)TiO$_3$ as depicted in Fig.~\ref{fig:half}(a).

In contrast to the parameters in effective Hamiltonian of Ref.~\onlinecite{Bellaiche.PhysRevB.73.144105}
obtained using LDA-based VCA, the present procedure gives improved estimation of equilibrium cubic lattice
constant $a_0$, and allow simple analysis of the effects of cationic disorder on ferroelectric transitions
(See Table~\ref{tab:parameters}).

\begin{table}
  \caption{Comparison of sets of parameters for BaTiO$_3$, SrTiO$_3$, and (Ba$_{x}$Sr$_{1-x}$)TiO$_3$ (BST).
    $p$ is the effective pressures applied during MD simulations.
    Details of these symbols are described in
    Refs.~\onlinecite{Nishimatsu:feram:PRB2008}~and~\onlinecite{Nishimatsu.PhysRevB.82.134106}.
  }
  \label{tab:parameters}
  \centering
  \begin{tabular}{rlrrrr}
    \hline
    \hline
             &               & Ref.~\onlinecite{Nishimatsu.PhysRevB.82.134106} \ %
                                  & \multicolumn{2}{c}{present work} & Ref.~\onlinecite{Bellaiche.PhysRevB.73.144105}\\
    \multicolumn{2}{c}{parameter}
                           & BaTiO$_3$ & SrTiO$_3$ & BST & VCA \\
    \hline
    $p$      & [GPa]       & $-0.005T$ & $0.0$  & \ $6.0(0.5-x)$  & $-5.2$ \\
    \hline
    $a_0$    & [\AA]       &  3.986     & 3.901 &   3.944 & 3.901 \\
    $B_{11}$  & [eV]        & 126.73     &131.33 & 129.03 & 129.96 \\
    $B_{12}$  & [eV]        &  41.76     &36.26 &   39.01 &  43.81 \\
    $B_{44}$  & [eV]        &  49.24     &41.30 &  45.27  &  46.94 \\
    \hline
    $c$      & [eV]        &            &      & $-0.279$&        \\
    \hline
    $B_{1xx}$ & [eV/\AA$^2$] &  $-185.35$ & $-102.09$ & $-143.72$& $-191.72$\\
    $B_{1yy}$ & [eV/\AA$^2$] &  $-3.2809$ & 0.5299 &   $-1.3755$&    $-3.98$\\
    $B_{4yz}$ & [eV/\AA$^2$] &  $-14.550$ & $-15.494$ &  $-15.022$&  $-5.73$\\
    $\alpha$ & [eV/\AA$^4$] &    78.99   & 22.39 & 50.69& 97.44\\
    $\gamma$ & [eV/\AA$^4$] & $-115.48$  & $-28.88$ & $-72.18$& $-143.25$ \\
    $k_1$    & [eV/\AA$^6$] & $-267.98$  & $-65.14$ &$-166.56$ & \\
    $k_2$    & [eV/\AA$^6$] &   197.50   & 117.00 & 157.25& \\
    $k_3$    & [eV/\AA$^6$] &   830.20   & 201.68 & 515.94& \\
    $k_4$    & [eV/\AA$^8$] &   641.97   & 139.35 & 390.66& \\
    \hline
    $M^*_\mathrm{dipole}$&[amu]&  38.24     & 43.61  & 40.93 & \\
    $M^*_\mathrm{acoustic}$&[amu]&46.64     & 36.70  & 41.67 & \\
    $Z^*$    & [e]          &  10.33     & 9.28 &  9.81& 9.66 \\
    $\epsilon_\infty$ &      &   6.87 & 6.46 & 6.66& 5.21 \\
    $\kappa_2$&[eV/\AA$^2$] &   8.534    & 10.316 &   9.425& 6.287 \\
    $j_1$    & [eV/\AA$^2$] & $-2.084$   &$-2.012$& $-2.048$& $-2.334$\\
    $j_2$    & [eV/\AA$^2$] & $-1.129$   &$-1.815$& $-1.472$& 4.318\\
    $j_3$    & [eV/\AA$^2$] &  0.689     &  0.590 &   0.640& 0.817\\
    $j_4$    & [eV/\AA$^2$] & $-0.611$   &$-0.567$& $-0.589$& $-0.461$ \\
    $j_5$    & [eV/\AA$^2$] &  0.000     &  0.000 &   0.000& 0.687\\
    $j_6$    & [eV/\AA$^2$] &  0.277     &  0.238 &   0.258& 0.147\\
    $j_7$    & [eV/\AA$^2$] &  0.000     &  0.000 &   0.000& 0.073\\
    \hline
    \hline
    $\kappa$&[eV/\AA$^2$] &   $-1.518$   & $-0.126$ & & \\
    \hline
    $\kappa(\Gamma_\mathrm{TO})$
            &[eV/\AA$^2$] &    $-1.906$  &$-0.254 $ & & \\
    $\kappa(\mathrm{X}_1)$
            &[eV/\AA$^2$] &     17.128   & 19.215 & & \\
    $\kappa(\mathrm{X}_5)$
            &[eV/\AA$^2$] &    $-1.422$  &  0.711 & & \\
    $\kappa(\mathrm{M}_{3'})$
            &[eV/\AA$^2$] &    $-1.143$  &  1.191 & & \\
    $\kappa(\mathrm{M}_{5'})$
            &[eV/\AA$^2$] &     16.333   &  18.424& & \\
    $\kappa(\mathrm{R}_{25'})$
            &[eV/\AA$^2$] &     13.871   & 16.300 & & \\
    \hline
    \hline
    $\xi^A_z$ &           &    0.166      &   0.4570 & & \\
    $\xi^B_z$ &           &    0.770      &   0.6302 & & \\
    $\xi^{\mathrm{O}_\mathrm{I}}  _z$ & & $-0.202$   &  $-0.3843$ & & \\
    $\xi^{\mathrm{O}_\mathrm{II}} _z$ & &  $-0.202$   & $-0.3843$ & & \\
    $\xi^{\mathrm{O}_\mathrm{III}}_z$ & &  $-0.546$   & $-0.3139$ & & \\
    \hline
    \hline
    $Z^{*A}_{zz}$ & [e]         & 2.741     &   2.565 & & \\
    $Z^{*B}_{zz}$ & [e]         & 7.492     &   7.435 & & \\
    $Z^{*\mathrm{O}_\mathrm{I}}  _{zz}$ &[e]&$-2.150$ &$-2.052$ & & \\
    $Z^{*\mathrm{O}_\mathrm{II}} _{zz}$ &[e]&$-2.150$ &$-2.052$  & & \\
    $Z^{*\mathrm{O}_\mathrm{III}}_{zz}$ &[e]&$-5.933$ &$-5.892$   & & \\
    \hline
    \hline
  \end{tabular}
\end{table}

\subsection{Determination of Parameters for (Ba$_{x}$Sr$_{1-x}$)TiO$_3$ Alloy}
\label{subsec:alloying}
To simulate an alloy or solid solution (Ba$_{x}$Sr$_{1-x}$)TiO$_3$,
we determine the modulation strength $c$ in Eq.~(\ref{eq:modulation}) as
\begin{equation}
  \label{eq:c}
  c = -\frac{1}{16}\frac{a_\mathrm{BTO}-a_\mathrm{STO}}{a_\mathrm{BST}}(B_{11}+2B_{12}) = -0.279~\mathrm{[eV]}\ .
  % gnuplot> print -1.0/16 * (3.986-3.901)/3.944 * (129.03 + 2*39.01) => -0.27889
\end{equation}
When $s(\bm{R})=\pm 8$, local inhomogeneous strains of
\begin{equation}
  \label{eq:minimizes}
  \eta_1=\eta_2=\eta_3=\pm \frac{1}{2}\frac{a_\mathrm{BTO}-a_\mathrm{STO}}{a_\mathrm{BST}}
  %  \eta_4=\eta_5=\eta_6&=0
\end{equation}
minimize the energy:
% of the local inhomogeneous strains at the $\bm{R}$ in the Hamiltonian
\begin{multline}
  \label{eq:Ee}
  E(\{\eta_\alpha\}_\mathrm{local})=\\
  \frac{1}{2}B_{11}(\eta_1^2 + \eta_2^2 + \eta_3^2) + B_{12}(\eta_2\eta_3 + \eta_3\eta_1 + \eta_1\eta_2)\\
  + \frac{1}{2}B_{44}(\eta_4^2 + \eta_5^2 + \eta_6^2) + cs(\eta_1 + \eta_2 + \eta_3)\ .
\end{multline}
Here,
$a_\mathrm{BTO}=3.986$~\AA\ is the calculated cubic lattice constant of BaTiO$_3$,
$a_\mathrm{STO}=3.901$~\AA\ is that of SrTiO$_3$, and
$a_\mathrm{BST}= (a_\mathrm{BTO} + a_\mathrm{STO})/2=3.944$~\AA\ ,
and $B_{11}$, $B_{12}$, and $B_{44}$ are the elastic constants of (Ba$_{1/2}$Sr$_{1/2}$)TiO$_3$ expressed in energy unit
($B_{11}=a_\mathrm{BST}^3C_{11}$, $B_{12}=a_\mathrm{BST}^3C_{12}$, and $B_{44}=a_\mathrm{BST}^3C_{44}$).
In Fig.~\ref{fig:half},
results of heating-up and cooling-down
MD simulations with four different values of $c=0$, $-0.279$, $-0.7$, and $-1.0$~[eV] are
given for $x=1/2$, i.e. (Ba$_{1/2}$Sr$_{1/2}$)TiO$_3$
for which the largest influence of disordered ionic configurations has to be expected.
We find lower transition temperature between
orthorhombic and rhombohedral structures, i.e. $\TOR$, for larger $|c|$.
The other two transition temperatures $\TCT$ and $\TTO$ remain almost unchanged.
This may be because $\TOR$ is the lowest transition temperature among the three,
and local inhomogeneous strains around each site
are almost frozen into the lowest energy structure.
However, the difference in $\TOR$ between $c=0$ and $c=-0.279$~eV is only 3~K.
% app26$ cat /home/t-nissie/feram/feram-p.08/xBST/xBST-04-32x32x32-b6.0-p0-c00.0000-vs/x0.50/Tc.dat
% 142.00 157.00  171.00 182.00  231.00 232.00
% app26$ cat /home/t-nissie/feram/feram-q.00/xBST/xBST-05-32x32x32-b6.0-p0-c-0.279/x0.5/Tc.dat
% 140.00 153.00   167.00 182.00   231.00 233.00
In Fig.~\ref{fig:half}(d),
we find strange behavior in tetragonal phase for $c=-1.0$
which may be unrealistically negatively large.

We first set the constant $b$ in the effective pressure of Eq.~(\ref{eq:effective:p})
so that average lattice constant becomes the same as BaTiO$_3$ for $x=1.0$ and SrTiO$_3$ for $x=0.0$ as
\begin{equation}
  \label{eq:b}
  b = \frac{a_\mathrm{BTO}-a_\mathrm{STO}}{a_\mathrm{BST}}3K = 11.65~\mathrm{[GPa]},
  % gnuplot> print (3.986-3.901)/3.944 * (129.03 + 2*39.01) * 1.60218e-19 / 3.944e-10**3 * 1.0e-9 => 11.653
\end{equation}
where $K=(B_{11}+2B_{12})/(3a_\mathrm{BST}^3)$ is bulk modulus.
However, $b=11.65$~[GPa] gives too high transition temperatures for $x=1$, i.e. BaTiO$_3$.
Therefore, we determine this $b$ empirically, as $b=6.0$~[GPa].
The reason for this may be the overestimation of the
coupling between homogeneous strain and polarization.

\subsection{Results of Molecular-Dynamics Simulations}
\label{subsec:MD}
\begin{figure}
  \centering
  \includegraphics[width=88mm]{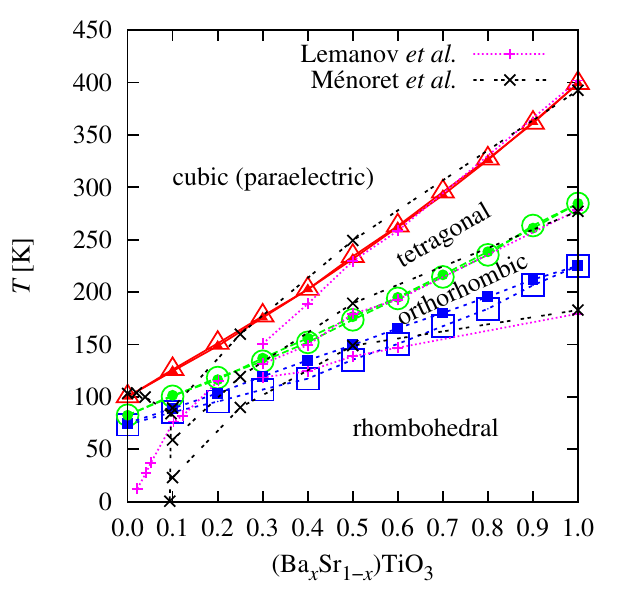}
  \caption{(Color online) Simulated temperature--composition ($T$--$x$) phase diagram.
    Heating-up and cooling-down transition temperatures are averaged
    when the corresponding transition has a hysteresis.
    Empty marks are from simulations with \textit{modulation} of $c=-0.7$~eV.
    Filled marks are from simulations without \textit{modulation} $c=0$.
    Note that results of $c=-0.7$~eV are shown here,
    because the difference between $c=0$ and $c=-0.279$~eV
    cannot be clearly seen in this scale.
    Two experimental results by
    Lemanov  \textit{et al.}\cite{PhysRevB.54.3151.BST.Lemanov} and
    M\'noret \textit{et al.}\cite{PhysRevB.65.224104}
    are also plotted for comparison.}
  \label{fig:phase:diagram}
  % /Users/takeshi/feram/feram-q.00/xBST/xBST-07-32x32x32-b6.0-p0-c-0.700/Tc-c-0_7.pdf
\end{figure}

Using the set of parameters determined above,
we perform heating-up and cooling-down MD simulations.
In Fig.~\ref{fig:phase:diagram},
a calculated temperature--composition ($T$--$x$) phase diagram is presented.
Heating-up and cooling-down transition temperatures are averaged
when corresponding transition has temperature hysteresis
between the heating-up and cooling-down simulations.

%\textbf{Anna's idea:}
For $x>0.25$, the almost linear $x$-dependence of all three transition temperatures is
well reproduced by our approach. For $x$ below $0.25$ the experimentally observed transition
temperatures decrease with a larger slope and
the alloy is no longer ferroelectric\cite{PhysRevB.65.224104} for $x < 0.094$.
In this concentration range of $x<0.25$,
the antiferrodistortive instability
found in pure SrTiO$_3$ may play an important role and
the instability reduces transition temperatures non-linearly,
and finally for pure SrTiO$_3$ the system is a quantum paraelectric.
Both effects are not accessible in our classical MD simulations neglecting rotations of octahedra.

\begin{figure}
  \centering
  \includegraphics[width=88mm]{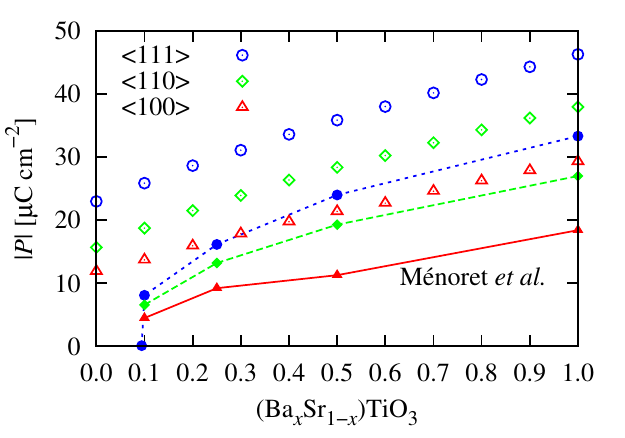}
  \caption{(Color online) Simulated composition ratio $x$-dependence of polarization $|P|$ (open marks).
    $|P|$ is measured at the middle of the two transition temperatures,
    $(\TCT+\TTO)/2$ for $|P_{\langle 100 \rangle}|$ of the tetragonal phase or
    $(\TTO+\TOR)/2$ for $|P_{\langle 110 \rangle}|$ of the orthorhombic phase,
    or half of $\TOR$ for $|P_{\langle 111 \rangle}|$ of the rhombohedral phase.
    For comparison, experimentally observed values
    by M\'noret \textit{et al.}\cite{PhysRevB.65.224104}
    are also plotted (filled marks connected with lines).}
  \label{fig:P}
  % ~/feram/feram-p.08/xBST/xBST-03-32x32x32-b6.0-p0-c-0.6554-vs/P-vs.pdf
\end{figure}

Simulated $x$-dependence of lattice constants $a$ and $c$ at room temperature (300~K)
is compared with experimental values\cite{McQuarrie:JACE:v38:p444:y1955} in Fig.~\ref{fig:P}.
Though the absolute values have good agreement,
more moderate $x$-dependence of lattice constants of
our simulations than the experiment is coming from
the empirical correction to $b$ from $11.65$ to $6.0$~GPa used here.
Overestimation of $c/a$ of this MD simulation is
coming from the error in first-principles calculations and
unavoidable within current techniques of DFT theories\cite{Sun:Umesh:Xifan:Perdew:NatChem:2016}.
\begin{figure}
  \centering
  \includegraphics[width=88mm]{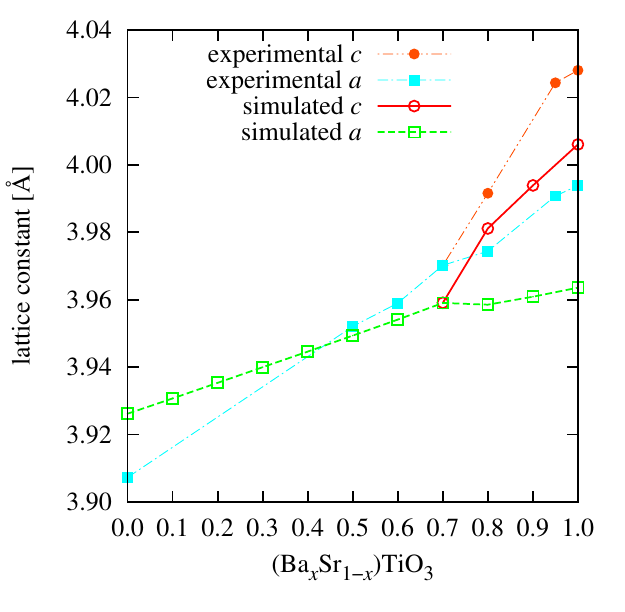}
  \caption{(Color online) Simulated $x$-dependence of lattice constants
    $a$ (open square marks) and
    $c$ (open circle marks) at room temperature (300~K).
    For comparison, experimentally observed values
    by McQuarrie\cite{McQuarrie:JACE:v38:p444:y1955} are also plotted (filled marks).}
  \label{fig:lattice}
  % ~/feram/feram-p.08/xBST/xBST-03-32x32x32-b6.0-p0-c-0.6554-vs/P-vs.pdf
\end{figure}

\begin{figure}
  \centering
  \includegraphics[width=65mm]{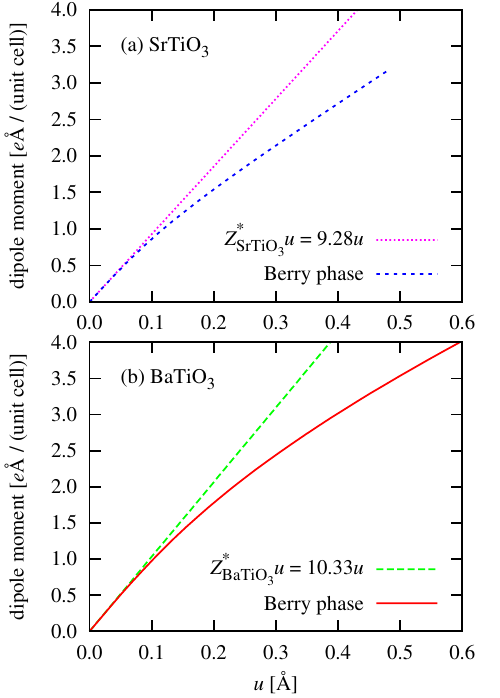}
  \caption{(Color online) Using the Berry-phase theory\cite{King-Smith:V:PRB:47:p1651-1654:1993},
    dipole moments per unit cell as a function of $u$
    for atomic displacements along $[001]$ distortion are calculated for (a)~SrTiO$_3$ and (b)~BaTiO$_3$.
    $Z^*u$ in $H^\mathrm{eff}$ of Eq.~(\ref{eq:Effective:Hamiltonian}) are also plotted for comparison.
    1.0~$e$\AA~in a unit cell of volume (4.0~\AA)$^3$ corresponds
    25~$\mu$C$\cdot$cm$^{-2}$.}
  \label{fig:berry}
  % appb:/home/t-nissie/abinit/BaTiO3-phonon/BaTiO3-phonon-67-WuCohenGGA-0.0GPa-berry/perovskite-001-appb2/BaTiO3-Berry-001.eps
\end{figure}

Simulated $x$-dependence of polarization $|P|$ is also compared
with the experimentally observed values\cite{PhysRevB.65.224104} in Fig.~\ref{fig:P}.
It is seen that our simulation slightly overestimates $|P|$ for
the whole range of $x$ and for any phases,
but trends for $x > 0.094$ are quite reasonable.
The main reason for this may come from the unavoidable overestimation of
$c/a$ and resulting overestimation of $|P|$ in first-principles calculations.
Moreover, as shown in Fig.~\ref{fig:berry},
%$u$-linearity of
\textit{true} dipole moment per unit cell $P(u)$ deviates from linearity at large $u$
both in BaTiO$_3$ and SrTiO$_3$,
and it may also explain the overestimation of $|P|$ in Fig.~\ref{fig:P}.
In Fig.~\ref{fig:berry}, \textit{true} dipole moment
as a function of $u$ for atomic displacements along $[001]$ distortion
calculated with the Berry-phase theory\cite{King-Smith:V:PRB:47:p1651-1654:1993}
is compared with $Z^*u$ in $H^\mathrm{eff}$ of Eq.~(\ref{eq:Effective:Hamiltonian}).

\section{Summary}
\label{sec:summary}
In this work, we presented a new set of parameters of an effective Hamiltonian
for (Ba$_{x}$Sr$_{1-x}$)TiO$_3$ solid-solution
%perovskite-type ferroelectric material
with input largely from the results of first-principles calculations,
including the \textit{modulation} in local inhomogeneous strains due to
Ba:Sr disorder. Using heating-up and cooling-down MD simulations
we have obtained $x-T$ phased diagram.
Though two parameters have been determined semi-empirically,
results of our simulations and experimentally observed values are in good agreement for
the dependence of transition temperatures, lattice constants $a$ and $c$,
and polarization on composition $x$.
It is found that $x$-dependent properties of (Ba$_{x}$Sr$_{1-x}$)TiO$_3$ are determined
mainly through the change in homogeneous lattice constants and that influence of
\textit{modulation} on local inhomogeneous strains is relatively weak.
%However, we expect effects of the latter to be notable when polarization domains
%and short-range order between Ba and Sr (e.g. in superlattices) are present {\bf REFER APL???}.

Our procedure of first-principles calculations, determination of parameters, and MD calculations
should be applicable to other perovskite-type ferroelectric solid solutions such as
(K,Na)NbO$_3$, (Ba,Sr,Ca)TiO$_3$, etc.

\section*{Acknowledgments}
Computational resources
were provided by the Center for Computational Materials Science,
Institute for Materials Research (CCMS-IMR), Tohoku University.
We thank the staff at CCMS-IMR for their constant effort.
This study is financially supported by the German Research Foundation, DFG SPP 1599.
This study was also supported in part by MEXT as a social and scientific
priority issue (Creation of new functional devices and high-performance
materials to support next-generation industries) to be tackled by using post-K computer.
U.V.W. acknowledges support from a JC Bose National Fellowship of the Department of Science and Technology,
Government of India.
We acknowledge collaboration and useful discussion with Anil Kumar.
% Bibliography Start
\bibliography{biblio/ferroelectrics,biblio/abinitio,biblio/MD}

%merlin.mbs apsrev4-1.bst 2010-07-25 4.21a (PWD, AO, DPC) hacked
%Control: key (0)
%Control: author (8) initials jnrlst
%Control: editor formatted (1) identically to author
%Control: production of article title (-1) disabled
%Control: page (0) single
%Control: year (1) truncated
%Control: production of eprint (0) enabled
\begin{thebibliography}{25}%
\makeatletter
\providecommand \@ifxundefined [1]{%
 \@ifx{#1\undefined}
}%
\providecommand \@ifnum [1]{%
 \ifnum #1\expandafter \@firstoftwo
 \else \expandafter \@secondoftwo
 \fi
}%
\providecommand \@ifx [1]{%
 \ifx #1\expandafter \@firstoftwo
 \else \expandafter \@secondoftwo
 \fi
}%
\providecommand \natexlab [1]{#1}%
\providecommand \enquote  [1]{``#1''}%
\providecommand \bibnamefont  [1]{#1}%
\providecommand \bibfnamefont [1]{#1}%
\providecommand \citenamefont [1]{#1}%
\providecommand \href@noop [0]{\@secondoftwo}%
\providecommand \href [0]{\begingroup \@sanitize@url \@href}%
\providecommand \@href[1]{\@@startlink{#1}\@@href}%
\providecommand \@@href[1]{\endgroup#1\@@endlink}%
\providecommand \@sanitize@url [0]{\catcode `\\12\catcode `\$12\catcode
  `\&12\catcode `\#12\catcode `\^12\catcode `\_12\catcode `\%12\relax}%
\providecommand \@@startlink[1]{}%
\providecommand \@@endlink[0]{}%
\providecommand \url  [0]{\begingroup\@sanitize@url \@url }%
\providecommand \@url [1]{\endgroup\@href {#1}{\urlprefix }}%
\providecommand \urlprefix  [0]{URL }%
\providecommand \Eprint [0]{\href }%
\providecommand \doibase [0]{http://dx.doi.org/}%
\providecommand \selectlanguage [0]{\@gobble}%
\providecommand \bibinfo  [0]{\@secondoftwo}%
\providecommand \bibfield  [0]{\@secondoftwo}%
\providecommand \translation [1]{[#1]}%
\providecommand \BibitemOpen [0]{}%
\providecommand \bibitemStop [0]{}%
\providecommand \bibitemNoStop [0]{.\EOS\space}%
\providecommand \EOS [0]{\spacefactor3000\relax}%
\providecommand \BibitemShut  [1]{\csname bibitem#1\endcsname}%
\let\auto@bib@innerbib\@empty
%</preamble>
\bibitem [{\citenamefont {Shannon}(1976)}]{ShannonEffectiveIonicRadii}%
  \BibitemOpen
  \bibfield  {author} {\bibinfo {author} {\bibfnamefont {R.~D.}\ \bibnamefont
  {Shannon}},\ }\href@noop {} {\bibfield  {journal} {\bibinfo  {journal} {Acta
  Cryst. A}\ }\textbf {\bibinfo {volume} {32}},\ \bibinfo {pages} {751}
  (\bibinfo {year} {1976})}\BibitemShut {NoStop}%
\bibitem [{\citenamefont {Goldschmidt}(1926)}]{Goldschmidt:ToleranceFactor}%
  \BibitemOpen
  \bibfield  {author} {\bibinfo {author} {\bibfnamefont {V.~M.}\ \bibnamefont
  {Goldschmidt}},\ }\href@noop {} {\bibfield  {journal} {\bibinfo  {journal}
  {Akad. Oslo Math-Natur.}\ }\textbf {\bibinfo {volume} {2}},\ \bibinfo {pages}
  {7} (\bibinfo {year} {1926})}\BibitemShut {NoStop}%
\bibitem [{\citenamefont {Fu}\ and\ \citenamefont
  {Itoh}(2011)}]{Fu:Itoh:Silver.Perovskite.Oxides:2011}%
  \BibitemOpen
  \bibfield  {author} {\bibinfo {author} {\bibfnamefont {D.}~\bibnamefont
  {Fu}}\ and\ \bibinfo {author} {\bibfnamefont {M.}~\bibnamefont {Itoh}},\
  }\href {\doibase 10.5772/17261} {\emph {\bibinfo {title} {Ferroelectrics --
  Material Aspects}}}\ (\bibinfo  {publisher} {INTECH},\ \bibinfo {address}
  {Rijeka},\ \bibinfo {year} {2011})\ \bibinfo {note} {chapter 20}\BibitemShut
  {NoStop}%
\bibitem [{\citenamefont {Fujishita}\ \emph {et~al.}(1979)\citenamefont
  {Fujishita}, \citenamefont {Shiozaki},\ and\ \citenamefont
  {Sawaguchi}}]{doi:10.1143/JPSJ.46.581}%
  \BibitemOpen
  \bibfield  {author} {\bibinfo {author} {\bibfnamefont {H.}~\bibnamefont
  {Fujishita}}, \bibinfo {author} {\bibfnamefont {Y.}~\bibnamefont {Shiozaki}},
  \ and\ \bibinfo {author} {\bibfnamefont {E.}~\bibnamefont {Sawaguchi}},\
  }\href {\doibase 10.1143/JPSJ.46.581} {\bibfield  {journal} {\bibinfo
  {journal} {J. Phys. Soc. Jpn.}\ }\textbf {\bibinfo {volume} {46}},\ \bibinfo
  {pages} {581} (\bibinfo {year} {1979})}\BibitemShut {NoStop}%
\bibitem [{\citenamefont {M\"uller}\ and\ \citenamefont
  {Burkard}(1979)}]{MULLER:B:PRB:19:p3593-3602:1979}%
  \BibitemOpen
  \bibfield  {author} {\bibinfo {author} {\bibfnamefont {K.~A.}\ \bibnamefont
  {M\"uller}}\ and\ \bibinfo {author} {\bibfnamefont {H.}~\bibnamefont
  {Burkard}},\ }\href@noop {} {\bibfield  {journal} {\bibinfo  {journal} {Phys.
  Rev. B}\ }\textbf {\bibinfo {volume} {19}},\ \bibinfo {pages} {3593}
  (\bibinfo {year} {1979})}\BibitemShut {NoStop}%
\bibitem [{\citenamefont {Lemanov}\ \emph {et~al.}(1996)\citenamefont
  {Lemanov}, \citenamefont {Smirnova}, \citenamefont {Syrnikov},\ and\
  \citenamefont {Tarakanov}}]{PhysRevB.54.3151.BST.Lemanov}%
  \BibitemOpen
  \bibfield  {author} {\bibinfo {author} {\bibfnamefont {V.~V.}\ \bibnamefont
  {Lemanov}}, \bibinfo {author} {\bibfnamefont {E.~P.}\ \bibnamefont
  {Smirnova}}, \bibinfo {author} {\bibfnamefont {P.~P.}\ \bibnamefont
  {Syrnikov}}, \ and\ \bibinfo {author} {\bibfnamefont {E.~A.}\ \bibnamefont
  {Tarakanov}},\ }\href {\doibase 10.1103/PhysRevB.54.3151} {\bibfield
  {journal} {\bibinfo  {journal} {Phys. Rev. B}\ }\textbf {\bibinfo {volume}
  {54}},\ \bibinfo {pages} {3151} (\bibinfo {year} {1996})}\BibitemShut
  {NoStop}%
\bibitem [{\citenamefont {M\'enoret}\ \emph {et~al.}(2002)\citenamefont
  {M\'enoret}, \citenamefont {Kiat}, \citenamefont {Dkhil}, \citenamefont
  {Dunlop}, \citenamefont {Dammak},\ and\ \citenamefont
  {Hernandez}}]{PhysRevB.65.224104}%
  \BibitemOpen
  \bibfield  {author} {\bibinfo {author} {\bibfnamefont {C.}~\bibnamefont
  {M\'enoret}}, \bibinfo {author} {\bibfnamefont {J.~M.}\ \bibnamefont {Kiat}},
  \bibinfo {author} {\bibfnamefont {B.}~\bibnamefont {Dkhil}}, \bibinfo
  {author} {\bibfnamefont {M.}~\bibnamefont {Dunlop}}, \bibinfo {author}
  {\bibfnamefont {H.}~\bibnamefont {Dammak}}, \ and\ \bibinfo {author}
  {\bibfnamefont {O.}~\bibnamefont {Hernandez}},\ }\href {\doibase
  10.1103/PhysRevB.65.224104} {\bibfield  {journal} {\bibinfo  {journal} {Phys.
  Rev. B}\ }\textbf {\bibinfo {volume} {65}},\ \bibinfo {pages} {224104}
  (\bibinfo {year} {2002})}\BibitemShut {NoStop}%
\bibitem [{\citenamefont {McQuarrie}(1955)}]{McQuarrie:JACE:v38:p444:y1955}%
  \BibitemOpen
  \bibfield  {author} {\bibinfo {author} {\bibfnamefont {M.}~\bibnamefont
  {McQuarrie}},\ }\href {\doibase 10.1111/j.1151-2916.1955.tb14571.x}
  {\bibfield  {journal} {\bibinfo  {journal} {J. Am. Ceram. Soc.}\ }\textbf
  {\bibinfo {volume} {38}},\ \bibinfo {pages} {444} (\bibinfo {year}
  {1955})}\BibitemShut {NoStop}%
\bibitem [{\citenamefont {Walizer}\ \emph {et~al.}(2006)\citenamefont
  {Walizer}, \citenamefont {Lisenkov},\ and\ \citenamefont
  {Bellaiche}}]{Bellaiche.PhysRevB.73.144105}%
  \BibitemOpen
  \bibfield  {author} {\bibinfo {author} {\bibfnamefont {L.}~\bibnamefont
  {Walizer}}, \bibinfo {author} {\bibfnamefont {S.}~\bibnamefont {Lisenkov}}, \
  and\ \bibinfo {author} {\bibfnamefont {L.}~\bibnamefont {Bellaiche}},\ }\href
  {\doibase 10.1103/PhysRevB.73.144105} {\bibfield  {journal} {\bibinfo
  {journal} {Phys. Rev. B}\ }\textbf {\bibinfo {volume} {73}},\ \bibinfo
  {pages} {144105} (\bibinfo {year} {2006})}\BibitemShut {NoStop}%
\bibitem [{\citenamefont {Gonze}\ \emph {et~al.}(2002)\citenamefont {Gonze},
  \citenamefont {Beuken}, \citenamefont {Caracas}, \citenamefont {Detraux},
  \citenamefont {Fuchs}, \citenamefont {Rignanese}, \citenamefont {Sindic},
  \citenamefont {Verstraete}, \citenamefont {Zerah}, \citenamefont {Jollet},
  \citenamefont {Torrent}, \citenamefont {Roy}, \citenamefont {Mikami},
  \citenamefont {Ghosez}, \citenamefont {Raty},\ and\ \citenamefont
  {Allan}}]{Gonze:ABINIT.ComputMaterSci:2002}%
  \BibitemOpen
  \bibfield  {author} {\bibinfo {author} {\bibfnamefont {X.}~\bibnamefont
  {Gonze}}, \bibinfo {author} {\bibfnamefont {J.-M.}\ \bibnamefont {Beuken}},
  \bibinfo {author} {\bibfnamefont {R.}~\bibnamefont {Caracas}}, \bibinfo
  {author} {\bibfnamefont {F.}~\bibnamefont {Detraux}}, \bibinfo {author}
  {\bibfnamefont {M.}~\bibnamefont {Fuchs}}, \bibinfo {author} {\bibfnamefont
  {G.-M.}\ \bibnamefont {Rignanese}}, \bibinfo {author} {\bibfnamefont
  {L.}~\bibnamefont {Sindic}}, \bibinfo {author} {\bibfnamefont
  {M.}~\bibnamefont {Verstraete}}, \bibinfo {author} {\bibfnamefont
  {G.}~\bibnamefont {Zerah}}, \bibinfo {author} {\bibfnamefont
  {F.}~\bibnamefont {Jollet}}, \bibinfo {author} {\bibfnamefont
  {M.}~\bibnamefont {Torrent}}, \bibinfo {author} {\bibfnamefont
  {A.}~\bibnamefont {Roy}}, \bibinfo {author} {\bibfnamefont {M.}~\bibnamefont
  {Mikami}}, \bibinfo {author} {\bibfnamefont {P.}~\bibnamefont {Ghosez}},
  \bibinfo {author} {\bibfnamefont {J.-Y.}\ \bibnamefont {Raty}}, \ and\
  \bibinfo {author} {\bibfnamefont {D.~C.}\ \bibnamefont {Allan}},\ }\href@noop
  {} {\bibfield  {journal} {\bibinfo  {journal} {Comput. Mater. Sci.}\ }\textbf
  {\bibinfo {volume} {25}},\ \bibinfo {pages} {478} (\bibinfo {year}
  {2002})}\BibitemShut {NoStop}%
\bibitem [{\citenamefont {Gonze}\ \emph {et~al.}(2009)\citenamefont {Gonze},
  \citenamefont {Amadon}, \citenamefont {Anglade}, \citenamefont {Beuken},
  \citenamefont {Bottin}, \citenamefont {Boulanger}, \citenamefont {Bruneval},
  \citenamefont {Caliste}, \citenamefont {Caracas}, \citenamefont {Cote},
  \citenamefont {Deutsch}, \citenamefont {Genovese}, \citenamefont {Ghosez},
  \citenamefont {Giantomassi}, \citenamefont {Goedecker}, \citenamefont
  {Hamann}, \citenamefont {Hermet}, \citenamefont {Jollet}, \citenamefont
  {Jomard}, \citenamefont {Leroux}, \citenamefont {Mancini}, \citenamefont
  {Mazevet}, \citenamefont {Oliveira}, \citenamefont {Onida}, \citenamefont
  {Pouillon}, \citenamefont {Rangel}, \citenamefont {Rignanese}, \citenamefont
  {Sangalli}, \citenamefont {Shaltaf}, \citenamefont {Torrent}, \citenamefont
  {Verstraete}, \citenamefont {Zerah},\ and\ \citenamefont
  {Zwanziger}}]{ABINIT_CPC_2009}%
  \BibitemOpen
  \bibfield  {author} {\bibinfo {author} {\bibfnamefont {X.}~\bibnamefont
  {Gonze}}, \bibinfo {author} {\bibfnamefont {B.}~\bibnamefont {Amadon}},
  \bibinfo {author} {\bibfnamefont {P.-M.}\ \bibnamefont {Anglade}}, \bibinfo
  {author} {\bibfnamefont {J.-M.}\ \bibnamefont {Beuken}}, \bibinfo {author}
  {\bibfnamefont {F.}~\bibnamefont {Bottin}}, \bibinfo {author} {\bibfnamefont
  {P.}~\bibnamefont {Boulanger}}, \bibinfo {author} {\bibfnamefont
  {F.}~\bibnamefont {Bruneval}}, \bibinfo {author} {\bibfnamefont
  {D.}~\bibnamefont {Caliste}}, \bibinfo {author} {\bibfnamefont
  {R.}~\bibnamefont {Caracas}}, \bibinfo {author} {\bibfnamefont
  {M.}~\bibnamefont {Cote}}, \bibinfo {author} {\bibfnamefont {T.}~\bibnamefont
  {Deutsch}}, \bibinfo {author} {\bibfnamefont {L.}~\bibnamefont {Genovese}},
  \bibinfo {author} {\bibfnamefont {P.}~\bibnamefont {Ghosez}}, \bibinfo
  {author} {\bibfnamefont {M.}~\bibnamefont {Giantomassi}}, \bibinfo {author}
  {\bibfnamefont {S.}~\bibnamefont {Goedecker}}, \bibinfo {author}
  {\bibfnamefont {D.~R.}\ \bibnamefont {Hamann}}, \bibinfo {author}
  {\bibfnamefont {P.}~\bibnamefont {Hermet}}, \bibinfo {author} {\bibfnamefont
  {F.}~\bibnamefont {Jollet}}, \bibinfo {author} {\bibfnamefont
  {G.}~\bibnamefont {Jomard}}, \bibinfo {author} {\bibfnamefont
  {S.}~\bibnamefont {Leroux}}, \bibinfo {author} {\bibfnamefont
  {M.}~\bibnamefont {Mancini}}, \bibinfo {author} {\bibfnamefont
  {S.}~\bibnamefont {Mazevet}}, \bibinfo {author} {\bibfnamefont {M.~J.~T.}\
  \bibnamefont {Oliveira}}, \bibinfo {author} {\bibfnamefont {G.}~\bibnamefont
  {Onida}}, \bibinfo {author} {\bibfnamefont {Y.}~\bibnamefont {Pouillon}},
  \bibinfo {author} {\bibfnamefont {T.}~\bibnamefont {Rangel}}, \bibinfo
  {author} {\bibfnamefont {G.-M.}\ \bibnamefont {Rignanese}}, \bibinfo {author}
  {\bibfnamefont {D.}~\bibnamefont {Sangalli}}, \bibinfo {author}
  {\bibfnamefont {R.}~\bibnamefont {Shaltaf}}, \bibinfo {author} {\bibfnamefont
  {M.}~\bibnamefont {Torrent}}, \bibinfo {author} {\bibfnamefont {M.~J.}\
  \bibnamefont {Verstraete}}, \bibinfo {author} {\bibfnamefont
  {G.}~\bibnamefont {Zerah}}, \ and\ \bibinfo {author} {\bibfnamefont {J.~W.}\
  \bibnamefont {Zwanziger}},\ }\href {\doibase 10.1016/j.cpc.2009.07.007}
  {\bibfield  {journal} {\bibinfo  {journal} {Comput. Phys. Commun.}\ }\textbf
  {\bibinfo {volume} {180}},\ \bibinfo {pages} {2582} (\bibinfo {year}
  {2009})}\BibitemShut {NoStop}%
\bibitem [{\citenamefont {Gonze}\ \emph {et~al.}(2016)\citenamefont {Gonze},
  \citenamefont {Jollet}, \citenamefont {Abreu~Araujo}, \citenamefont {Adams},
  \citenamefont {Amadon}, \citenamefont {Applencourt}, \citenamefont {Audouze},
  \citenamefont {Beuken}, \citenamefont {Bieder}, \citenamefont {Bokhanchuk},
  \citenamefont {Bousquet}, \citenamefont {Bruneval}, \citenamefont {Caliste},
  \citenamefont {C\^ot\'e}, \citenamefont {Dahm}, \citenamefont {Da~Pieve},
  \citenamefont {Delaveau}, \citenamefont {Di~Gennaro}, \citenamefont {Dorado},
  \citenamefont {Espejo}, \citenamefont {Geneste}, \citenamefont {Genovese},
  \citenamefont {Gerossier}, \citenamefont {Giantomassi}, \citenamefont
  {Gillet}, \citenamefont {Hamann}, \citenamefont {He}, \citenamefont {Jomard},
  \citenamefont {Laflamme~Janssen}, \citenamefont {Le~Roux}, \citenamefont
  {Levitt}, \citenamefont {Lherbier}, \citenamefont {Liu}, \citenamefont
  {Luka\v{c}evi\'c}, \citenamefont {Martin}, \citenamefont {Martins},
  \citenamefont {Oliveira}, \citenamefont {Ponc\'e}, \citenamefont {Pouillon},
  \citenamefont {Rangel}, \citenamefont {Rignanese}, \citenamefont {Romero},
  \citenamefont {Rousseau}, \citenamefont {Rubel}, \citenamefont {Shukri},
  \citenamefont {Stankovski}, \citenamefont {Torrent}, \citenamefont
  {Van~Setten}, \citenamefont {Van~Troeye}, \citenamefont {Verstraete},
  \citenamefont {Waroquiers}, \citenamefont {Wiktor}, \citenamefont {Xu},
  \citenamefont {Zhou},\ and\ \citenamefont {Zwanziger}}]{ABINIT_CPC_2016}%
  \BibitemOpen
  \bibfield  {author} {\bibinfo {author} {\bibfnamefont {X.}~\bibnamefont
  {Gonze}}, \bibinfo {author} {\bibfnamefont {F.}~\bibnamefont {Jollet}},
  \bibinfo {author} {\bibfnamefont {F.}~\bibnamefont {Abreu~Araujo}}, \bibinfo
  {author} {\bibfnamefont {D.}~\bibnamefont {Adams}}, \bibinfo {author}
  {\bibfnamefont {B.}~\bibnamefont {Amadon}}, \bibinfo {author} {\bibfnamefont
  {T.}~\bibnamefont {Applencourt}}, \bibinfo {author} {\bibfnamefont
  {C.}~\bibnamefont {Audouze}}, \bibinfo {author} {\bibfnamefont {J.-M.}\
  \bibnamefont {Beuken}}, \bibinfo {author} {\bibfnamefont {J.}~\bibnamefont
  {Bieder}}, \bibinfo {author} {\bibfnamefont {A.}~\bibnamefont {Bokhanchuk}},
  \bibinfo {author} {\bibfnamefont {E.}~\bibnamefont {Bousquet}}, \bibinfo
  {author} {\bibfnamefont {F.}~\bibnamefont {Bruneval}}, \bibinfo {author}
  {\bibfnamefont {D.}~\bibnamefont {Caliste}}, \bibinfo {author} {\bibfnamefont
  {M.}~\bibnamefont {C\^ot\'e}}, \bibinfo {author} {\bibfnamefont
  {F.}~\bibnamefont {Dahm}}, \bibinfo {author} {\bibfnamefont {F.}~\bibnamefont
  {Da~Pieve}}, \bibinfo {author} {\bibfnamefont {M.}~\bibnamefont {Delaveau}},
  \bibinfo {author} {\bibfnamefont {M.}~\bibnamefont {Di~Gennaro}}, \bibinfo
  {author} {\bibfnamefont {B.}~\bibnamefont {Dorado}}, \bibinfo {author}
  {\bibfnamefont {C.}~\bibnamefont {Espejo}}, \bibinfo {author} {\bibfnamefont
  {G.}~\bibnamefont {Geneste}}, \bibinfo {author} {\bibfnamefont
  {L.}~\bibnamefont {Genovese}}, \bibinfo {author} {\bibfnamefont
  {A.}~\bibnamefont {Gerossier}}, \bibinfo {author} {\bibfnamefont
  {M.}~\bibnamefont {Giantomassi}}, \bibinfo {author} {\bibfnamefont
  {Y.}~\bibnamefont {Gillet}}, \bibinfo {author} {\bibfnamefont {D.~R.}\
  \bibnamefont {Hamann}}, \bibinfo {author} {\bibfnamefont {L.}~\bibnamefont
  {He}}, \bibinfo {author} {\bibfnamefont {G.}~\bibnamefont {Jomard}}, \bibinfo
  {author} {\bibfnamefont {J.}~\bibnamefont {Laflamme~Janssen}}, \bibinfo
  {author} {\bibfnamefont {S.}~\bibnamefont {Le~Roux}}, \bibinfo {author}
  {\bibfnamefont {A.}~\bibnamefont {Levitt}}, \bibinfo {author} {\bibfnamefont
  {A.}~\bibnamefont {Lherbier}}, \bibinfo {author} {\bibfnamefont
  {F.}~\bibnamefont {Liu}}, \bibinfo {author} {\bibfnamefont {I.}~\bibnamefont
  {Luka\v{c}evi\'c}}, \bibinfo {author} {\bibfnamefont {A.}~\bibnamefont
  {Martin}}, \bibinfo {author} {\bibfnamefont {C.}~\bibnamefont {Martins}},
  \bibinfo {author} {\bibfnamefont {M.~J.~T.}\ \bibnamefont {Oliveira}},
  \bibinfo {author} {\bibfnamefont {S.}~\bibnamefont {Ponc\'e}}, \bibinfo
  {author} {\bibfnamefont {Y.}~\bibnamefont {Pouillon}}, \bibinfo {author}
  {\bibfnamefont {T.}~\bibnamefont {Rangel}}, \bibinfo {author} {\bibfnamefont
  {G.-M.}\ \bibnamefont {Rignanese}}, \bibinfo {author} {\bibfnamefont {A.~H.}\
  \bibnamefont {Romero}}, \bibinfo {author} {\bibfnamefont {B.}~\bibnamefont
  {Rousseau}}, \bibinfo {author} {\bibfnamefont {O.}~\bibnamefont {Rubel}},
  \bibinfo {author} {\bibfnamefont {A.~A.}\ \bibnamefont {Shukri}}, \bibinfo
  {author} {\bibfnamefont {M.}~\bibnamefont {Stankovski}}, \bibinfo {author}
  {\bibfnamefont {M.}~\bibnamefont {Torrent}}, \bibinfo {author} {\bibfnamefont
  {M.~J.}\ \bibnamefont {Van~Setten}}, \bibinfo {author} {\bibfnamefont
  {B.}~\bibnamefont {Van~Troeye}}, \bibinfo {author} {\bibfnamefont {M.~J.}\
  \bibnamefont {Verstraete}}, \bibinfo {author} {\bibfnamefont
  {D.}~\bibnamefont {Waroquiers}}, \bibinfo {author} {\bibfnamefont
  {J.}~\bibnamefont {Wiktor}}, \bibinfo {author} {\bibfnamefont
  {B.}~\bibnamefont {Xu}}, \bibinfo {author} {\bibfnamefont {A.}~\bibnamefont
  {Zhou}}, \ and\ \bibinfo {author} {\bibfnamefont {J.~W.}\ \bibnamefont
  {Zwanziger}},\ }\href {\doibase doi:10.1016/j.cpc.2016.04.003} {\bibfield
  {journal} {\bibinfo  {journal} {Comput. Phys. Commun.}\ }\textbf {\bibinfo
  {volume} {205}},\ \bibinfo {pages} {106} (\bibinfo {year}
  {2016})}\BibitemShut {NoStop}%
\bibitem [{\citenamefont {Wu}\ and\ \citenamefont
  {Cohen}(2006)}]{Wu:C:PRB:73:p235116:2006}%
  \BibitemOpen
  \bibfield  {author} {\bibinfo {author} {\bibfnamefont {Z.~G.}\ \bibnamefont
  {Wu}}\ and\ \bibinfo {author} {\bibfnamefont {R.~E.}\ \bibnamefont {Cohen}},\
  }\href {\doibase 10.1103/PhysRevB.73.235116} {\bibfield  {journal} {\bibinfo
  {journal} {Phys. Rev. B}\ }\textbf {\bibinfo {volume} {73}},\ \bibinfo
  {pages} {235116} (\bibinfo {year} {2006})}\BibitemShut {NoStop}%
\bibitem [{\citenamefont {Rappe}\ \emph {et~al.}(1990)\citenamefont {Rappe},
  \citenamefont {Rabe}, \citenamefont {Kaxiras},\ and\ \citenamefont
  {Joannopoulos}}]{RAPPE:R:K:J:PRB:41:p1227-1230:1990}%
  \BibitemOpen
  \bibfield  {author} {\bibinfo {author} {\bibfnamefont {A.~M.}\ \bibnamefont
  {Rappe}}, \bibinfo {author} {\bibfnamefont {K.~M.}\ \bibnamefont {Rabe}},
  \bibinfo {author} {\bibfnamefont {E.}~\bibnamefont {Kaxiras}}, \ and\
  \bibinfo {author} {\bibfnamefont {J.~D.}\ \bibnamefont {Joannopoulos}},\
  }\href@noop {} {\bibfield  {journal} {\bibinfo  {journal} {Phys. Rev. B}\
  }\textbf {\bibinfo {volume} {41}},\ \bibinfo {pages} {1227} (\bibinfo {year}
  {1990})}\BibitemShut {NoStop}%
\bibitem [{opi(2014)}]{opium}%
  \BibitemOpen
  \href@noop {} {\emph {\bibinfo {title} {Opium - pseudopotential generation
  project, {\tt http://opium.sourceforge.net/}}}} (\bibinfo {year}
  {1998--2014})\BibitemShut {NoStop}%
\bibitem [{\citenamefont {Hashimoto}\ \emph {et~al.}(2004)\citenamefont
  {Hashimoto}, \citenamefont {Nishimatsu}, \citenamefont {Mizuseki},
  \citenamefont {Kawazoe}, \citenamefont {Sasaki},\ and\ \citenamefont
  {Ikeda}}]{Hashimoto:N:M:K:S:I:JJAP:43:p6785-6792:2004}%
  \BibitemOpen
  \bibfield  {author} {\bibinfo {author} {\bibfnamefont {T.}~\bibnamefont
  {Hashimoto}}, \bibinfo {author} {\bibfnamefont {T.}~\bibnamefont
  {Nishimatsu}}, \bibinfo {author} {\bibfnamefont {H.}~\bibnamefont
  {Mizuseki}}, \bibinfo {author} {\bibfnamefont {Y.}~\bibnamefont {Kawazoe}},
  \bibinfo {author} {\bibfnamefont {A.}~\bibnamefont {Sasaki}}, \ and\ \bibinfo
  {author} {\bibfnamefont {Y.}~\bibnamefont {Ikeda}},\ }\href@noop {}
  {\bibfield  {journal} {\bibinfo  {journal} {Jpn. J. Appl. Phys.}\ }\textbf
  {\bibinfo {volume} {43}},\ \bibinfo {pages} {6785} (\bibinfo {year}
  {2004})}\BibitemShut {NoStop}%
\bibitem [{\citenamefont {Nishimatsu}\ \emph {et~al.}(2010)\citenamefont
  {Nishimatsu}, \citenamefont {Iwamoto}, \citenamefont {Kawazoe},\ and\
  \citenamefont {Waghmare}}]{Nishimatsu.PhysRevB.82.134106}%
  \BibitemOpen
  \bibfield  {author} {\bibinfo {author} {\bibfnamefont {T.}~\bibnamefont
  {Nishimatsu}}, \bibinfo {author} {\bibfnamefont {M.}~\bibnamefont {Iwamoto}},
  \bibinfo {author} {\bibfnamefont {Y.}~\bibnamefont {Kawazoe}}, \ and\
  \bibinfo {author} {\bibfnamefont {U.~V.}\ \bibnamefont {Waghmare}},\ }\href
  {\doibase 10.1103/PhysRevB.82.134106} {\bibfield  {journal} {\bibinfo
  {journal} {Phys. Rev. B}\ }\textbf {\bibinfo {volume} {82}},\ \bibinfo
  {pages} {134106} (\bibinfo {year} {2010})}\BibitemShut {NoStop}%
\bibitem [{\citenamefont {Nishimatsu}\ \emph {et~al.}(2008)\citenamefont
  {Nishimatsu}, \citenamefont {Waghmare}, \citenamefont {Kawazoe},\ and\
  \citenamefont {Vanderbilt}}]{Nishimatsu:feram:PRB2008}%
  \BibitemOpen
  \bibfield  {author} {\bibinfo {author} {\bibfnamefont {T.}~\bibnamefont
  {Nishimatsu}}, \bibinfo {author} {\bibfnamefont {U.~V.}\ \bibnamefont
  {Waghmare}}, \bibinfo {author} {\bibfnamefont {Y.}~\bibnamefont {Kawazoe}}, \
  and\ \bibinfo {author} {\bibfnamefont {D.}~\bibnamefont {Vanderbilt}},\
  }\href@noop {} {\bibfield  {journal} {\bibinfo  {journal} {Phys. Rev. B}\
  }\textbf {\bibinfo {volume} {78}},\ \bibinfo {pages} {104104} (\bibinfo
  {year} {2008})}\BibitemShut {NoStop}%
\bibitem [{\citenamefont {King-Smith}\ and\ \citenamefont
  {Vanderbilt}(1994)}]{King-Smith:V:1994}%
  \BibitemOpen
  \bibfield  {author} {\bibinfo {author} {\bibfnamefont {R.~D.}\ \bibnamefont
  {King-Smith}}\ and\ \bibinfo {author} {\bibfnamefont {D.}~\bibnamefont
  {Vanderbilt}},\ }\href@noop {} {\bibfield  {journal} {\bibinfo  {journal}
  {Phys. Rev. B}\ }\textbf {\bibinfo {volume} {49}},\ \bibinfo {pages} {5828}
  (\bibinfo {year} {1994})}\BibitemShut {NoStop}%
\bibitem [{\citenamefont {Zhong}\ \emph {et~al.}(1995)\citenamefont {Zhong},
  \citenamefont {Vanderbilt},\ and\ \citenamefont
  {Rabe}}]{Zhong:V:R:PRB:v52:p6301:1995}%
  \BibitemOpen
  \bibfield  {author} {\bibinfo {author} {\bibfnamefont {W.}~\bibnamefont
  {Zhong}}, \bibinfo {author} {\bibfnamefont {D.}~\bibnamefont {Vanderbilt}}, \
  and\ \bibinfo {author} {\bibfnamefont {K.~M.}\ \bibnamefont {Rabe}},\
  }\href@noop {} {\bibfield  {journal} {\bibinfo  {journal} {Phys. Rev. B}\
  }\textbf {\bibinfo {volume} {52}},\ \bibinfo {pages} {6301} (\bibinfo {year}
  {1995})}\BibitemShut {NoStop}%
\bibitem [{\citenamefont {Waghmare}\ \emph {et~al.}(2003)\citenamefont
  {Waghmare}, \citenamefont {Cockayne},\ and\ \citenamefont
  {Burton}}]{Waghmare:C:B:2003}%
  \BibitemOpen
  \bibfield  {author} {\bibinfo {author} {\bibfnamefont {U.~V.}\ \bibnamefont
  {Waghmare}}, \bibinfo {author} {\bibfnamefont {E.~J.}\ \bibnamefont
  {Cockayne}}, \ and\ \bibinfo {author} {\bibfnamefont {B.~P.}\ \bibnamefont
  {Burton}},\ }\href@noop {} {\bibfield  {journal} {\bibinfo  {journal}
  {Ferroelectrics}\ }\textbf {\bibinfo {volume} {291}},\ \bibinfo {pages} {187}
  (\bibinfo {year} {2003})}\BibitemShut {NoStop}%
\bibitem [{\citenamefont {Waghmare}(2014)}]{ACR:Waghmare:2014}%
  \BibitemOpen
  \bibfield  {author} {\bibinfo {author} {\bibfnamefont {U.~V.}\ \bibnamefont
  {Waghmare}},\ }\href {\doibase 10.1021/ar500331c} {\bibfield  {journal}
  {\bibinfo  {journal} {Acc. Chem. Res.}\ }\textbf {\bibinfo {volume} {47}},\
  \bibinfo {pages} {3242} (\bibinfo {year} {2014})}\BibitemShut {NoStop}%
\bibitem [{\citenamefont {Nishimatsu}(2016)}]{feram}%
  \BibitemOpen
  \bibfield  {author} {\bibinfo {author} {\bibfnamefont {T.}~\bibnamefont
  {Nishimatsu}},\ }\href@noop {} {\emph {\bibinfo {title} {feram at
  SourceForge.net, {\tt http://loto.sourceforge.net/feram/}}}} (\bibinfo {year}
  {2007--2016})\BibitemShut {NoStop}%
\bibitem [{\citenamefont {Sun}\ \emph {et~al.}(2016)\citenamefont {Sun},
  \citenamefont {Remsing}, \citenamefont {Zhang}, \citenamefont {Sun},
  \citenamefont {Ruzsinszky}, \citenamefont {Peng}, \citenamefont {Yang},
  \citenamefont {Paul}, \citenamefont {Waghmare}, \citenamefont {Wu},
  \citenamefont {Klein},\ and\ \citenamefont
  {Perdew}}]{Sun:Umesh:Xifan:Perdew:NatChem:2016}%
  \BibitemOpen
  \bibfield  {author} {\bibinfo {author} {\bibfnamefont {J.}~\bibnamefont
  {Sun}}, \bibinfo {author} {\bibfnamefont {R.~C.}\ \bibnamefont {Remsing}},
  \bibinfo {author} {\bibfnamefont {Y.}~\bibnamefont {Zhang}}, \bibinfo
  {author} {\bibfnamefont {Z.}~\bibnamefont {Sun}}, \bibinfo {author}
  {\bibfnamefont {A.}~\bibnamefont {Ruzsinszky}}, \bibinfo {author}
  {\bibfnamefont {H.}~\bibnamefont {Peng}}, \bibinfo {author} {\bibfnamefont
  {Z.}~\bibnamefont {Yang}}, \bibinfo {author} {\bibfnamefont {A.}~\bibnamefont
  {Paul}}, \bibinfo {author} {\bibfnamefont {U.}~\bibnamefont {Waghmare}},
  \bibinfo {author} {\bibfnamefont {X.}~\bibnamefont {Wu}}, \bibinfo {author}
  {\bibfnamefont {M.~L.}\ \bibnamefont {Klein}}, \ and\ \bibinfo {author}
  {\bibfnamefont {J.~P.}\ \bibnamefont {Perdew}},\ }\href@noop {} {\bibfield
  {journal} {\bibinfo  {journal} {Nat. Chem.}\ ,\ \bibinfo {pages} {2535}}
  (\bibinfo {year} {2016})}\BibitemShut {NoStop}%
\bibitem [{\citenamefont {King-Smith}\ and\ \citenamefont
  {Vanderbilt}(1993)}]{King-Smith:V:PRB:47:p1651-1654:1993}%
  \BibitemOpen
  \bibfield  {author} {\bibinfo {author} {\bibfnamefont {R.~D.}\ \bibnamefont
  {King-Smith}}\ and\ \bibinfo {author} {\bibfnamefont {D.}~\bibnamefont
  {Vanderbilt}},\ }\href@noop {} {\bibfield  {journal} {\bibinfo  {journal}
  {Phys. Rev. B}\ }\textbf {\bibinfo {volume} {47}},\ \bibinfo {pages} {1651}
  (\bibinfo {year} {1993})}\BibitemShut {NoStop}%
\end{thebibliography}%
\end{document}